# Triple-helical collagen hydrogels via covalent aromatic functionalization with 1,3-Phenylenediacetic acid

Giuseppe Tronci,[1,2] Amanda Doyle,[1,2] Stephen J. Russell,[2] and David J. Wood[1]

[1] Biomaterials and Tissue Engineering Research Group, Leeds Dental Institute, University of Leeds, Leeds LS2 9LU, United Kingdom

[2] Nonwoven Research Group, Centre for Technical Textiles, University of Leeds, Leeds LS2 9JT, United Kingdom

## Table of contents entry

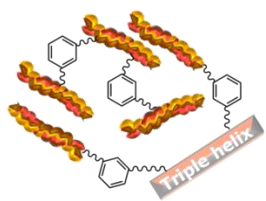 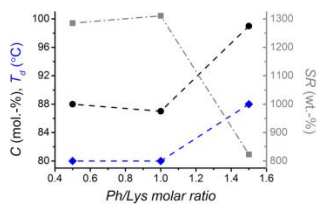

Covalent functionalization of type I collagen with 1,3-Phenylenediacetic acid (Ph) directly leads to triple-helical hydrogels with controlled degree of crosslinking ($C$), swelling ratio ($SR$) and denaturation temperature ($T_d$), in contrast to state-of-the-art carbodiimide crosslinking.

## Abstract

Chemical crosslinking of collagen is a general strategy to reproduce macroscale tissue properties in physiological environment. However, simultaneous control of protein conformation, material properties and biofunctionality is highly challenging with current synthetic strategies. Consequently, the potentially-diverse clinical applications of collagen-based biomaterials cannot be fully realised. In order to establish defined biomacromolecular systems for mineralised tissue applications, type I collagen was functionalised with 1,3-Phenylenediacetic acid (Ph) and investigated at the molecular, macroscopic and functional levels. Preserved triple helix conformation was observed in obtained covalent networks via ATR-FTIR ($A_{III}/A_{1450}$ ~ 1) and WAXS, while network crosslinking degree (*C*: 87-99 mol.-%) could be adjusted based on specific reaction conditions. Decreased swelling ratio (*SR*: 823-1285 wt.-%) and increased thermo-mechanical ($T_d$: 80-88 °C; *E*: 28-35 kPa; $\sigma_{max}$: 6-8 kPa; $\varepsilon_b$: 53-58 %) properties were observed compared to state-of-the-art carbodiimide (EDC)-crosslinked collagen controls, likely related to the intermolecular covalent incorporation of the aromatic segment. Ph-crosslinked hydrogels displayed nearly intact material integrity and only a slight mass decrease ($M_R$: 5-11 wt. %) following 1-week incubation in either PBS or simulated body fluid (SBF), in contrast to EDC-crosslinked collagen ($M_R$: 33-58 wt. %). Furthermore, FTIR, SEM and EDS revealed deposition of a calcium-phosphate phase on SBF-retrieved samples, whereby an increased calcium phosphate ratio (*Ca/P*: 0.84-1.41) was observed in hydrogels with higher Ph content. 72-hour material extracts were well tolerated by L929 mouse fibroblasts, whereby cell confluence and metabolic activity (MTS assay) were comparable to those of cells cultured in cell culture medium (positive control). In light of their controlled structure-function properties, these biocompatible collagen hydrogels represent attractive material systems for potential mineralised tissue applications.

## 1. Introduction

Biomaterials play a crucial role in regenerative medicine and pharmaceutics and find huge applications in wound care, orthopaedics and cardiovascular industries, among others.[1] Particularly for mineralised tissue applications, the design of versatile biomimetic systems, which can be safely implanted *in vivo*, displaying defined biodegradability, tuneable mechanical properties, and bioactivity, is a pressing, unmet, clinical need towards next generation therapeutics.[2] *In vivo*, such systems should provide a biomimetic interface to cells and induce desired biological processes, such as localised recruitment and controlled differentiation of stem cells,[3] as well as stimulating the formation of new bone via natural biomineralisation.[4] Although synthetic biodegradable polymers, e.g. polyesters, can be precisely tuned as to their chemical composition and material properties,[5] they are usually biologically inert and lack biofunctionality. Furthermore, the design of defined chemical systems based on building blocks derived from the organic matrix of tissues represents an interesting approach for the establishment of bespoke biomimetic materials with tissue-like architecture and composition.[6]

Collagen is the main protein of the human body, ruling structure, function and shape of biological tissues, such as bone.[7] Also in light of its unique molecular organisation, collagen has been widely applied for the design of vascular grafts,[8] fibrous materials for stem cell differentiation,[9] biomimetic scaffolds for regenerative medicine,[10] and tissue-like matrices for bone tissue repair.[11] However, collagen properties are challenging to control in physiological conditions, due to the fact that collagen's unique hierarchical organisation and chemical composition *in vivo* can only be partially reproduced *in vitro*. In its monomeric form, the collagen molecule is based on three left-handed polyproline chains, each one containing the repeating unit Gly-*X*-*Y*, where *X* and *Y* are predominantly proline (Pro) and hydroxyproline (Hyp), respectively. The three chains are staggered to one another by one amino acid residue and are twisted together to form a right-handed triple helix (300 nm in length, 1.5 nm in diameter). *In vivo*, triple helices can aggregate in a periodic staggered array to form collagen fibrils, fibres and fascicles; such structural hierarchies are stabilised via covalent crosslinking[12] and can be mineralised via apatite deposition to form new bone. Here, matrix hierarchical organisation directly impacts on the distribution and size of nucleated apatite crystals, so that a clear relationship between internal collagen architecture and bone physiological state is present.[13]

Fibrillogenesis can be induced *in vitro* by exposing triple helical collagen to physiological conditions. However, hydrogen and covalent bonds stabilising collagen structures *in vivo* are partially broken following isolation *ex-vivo*, so that collagen hierarchical organisation is affected. As a result, collagen materials display non-controllable swelling and weak mechanical properties in physiological conditions; reliable synthetic methods must therefore be applied in order to improve hydrogel thermo-mechanical behaviour, without affecting collagen biofunctionality,[14] i.e. biocompatibility and bioactivity. Following a bottom-up synthetic approach, O'Leary et al. proposed a self-assembling peptide system capable to recapitulate the triple helical and fibrillar architecture of collagen.[15] Based on the peptide concentration and peptide sequence, synthetic collagen-like hydrogels were successfully developed, although the thermo-mechanical stability was still not adequate for biomaterial applications. Exploiting the fibrillogenesis process of native collagen, Nam et al. proposed a heterostructural collagen matrix obtained via layer-by-layer deposition of collagen solutions at different densities.[16] By adjusting the collagen concentration and the number of layers, mechanical properties and cell proliferation could be controlled, although the degradation behaviour of these materials in physiological conditions was not investigated. Collagen fibrillogenesis was also exploited to replicate the hierarchical organization of mineralized tissues.[17] Here, polyacrylic acid was applied to stabilise an amorphous calcium phosphate phase during the self-assembly of collagen triple helices into fibrils. In this way, highly-ordered mineralized collagen matrices with improved mechanical properties were successfully formed, although the material behaviour in physiological conditions was not fully addressed. Recently, dynamic perfusion-flow mineralization techniques have also been proposed to recapitulate the natural mineralization process in collagen templates.[18]

Other than exploiting the inherent triple helix self-assembly, functionalisation and covalent crosslinking, e.g. via N-(3-Dimethylaminopropyl)-N′-ethylcarbodiimide hydrochloride (EDC),[19,20] glutaraldehyde,[21,22] hexamethylene diisocyanate[23] and enzymatic treatment[24] have been shown to enhance macroscopic properties and resistance to enzymatic degradation of collagen in biological environment.[25] However, safe and systematic control in network architecture has proved to be challenging, whereby occurrence of side reactions may lead to the formation of potentially-toxic materials with only slight variation in macroscopic properties. Diimidoesters-dimethyl suberimidate (DMS), 3,3'-dithiobispropionimidate (DTBP) and acyl azide have been employed as alternative crosslinking agents, resulting in stable materials in physiological conditions, although reduced material elasticity was observed.[26] Additionally, both dehydrothermal and riboflavin-mediated treatments have been applied as physical, benign stabilisation methods, leading to partial loss of native collagen structure and non-homogeneous crosslinking.[27] In an effort to design water-stable composite materials, collagen fibrils were functionalized with a silane moiety, in order to promote covalent links between the inorganic phase and the fibrous matrix.[28] The resulting materials displayed preserved collagen fibrous characteristics and enhanced degradability, although the mechanical properties were not fully investigated. From all the aforementioned examples it appears rather clear that while the macroscopic properties may be improved, systematic control of material behaviour at different length scales is only minimally accomplished in the resulting collagen materials. Thus, the establishment of novel synthetic methods ensuring tunable collagen functionalisation, preserved protein conformation together with full biocompatibility and bioactivity is necessary in order to establish reliable collagen systems with defined structure-function properties for mineralised tissue repair.

In this work, the design of type I collagen hydrogels via covalent lysine functionalisation with 1,3-Phenylenediacetic acid (Ph) was investigated. Given that the collagen displays cell-binding peptides at the molecular level, we investigated whether the incorporation of covalent crosslinking of collagen triple helices via a stiff, aromatic segment could offer a synthetic strategy to the formation of defined biomacromolecular materials with relevant mechanical properties and retained biochemical features. Ph was selected as a novel bifunctional segment, in order to promote inter-molecular crosslinking of collagen molecules, unlikely to be accomplished with current synthetic methods.[8,19-23] This specific functionalisation was aimed to achieve controlled swelling and enhanced mechanical properties in resulting hydrogels, due to the backbone rigidity and hydrophobicity of the Ph aromatic segment. In order to investigate the effectiveness of this synthetic approach over current strategies, EDC treatment was selected as a state-of-the-art reference method, since it has been shown to promote the formation of water-stable collagen materials with no residual toxicity,[25] in contrast to aldehyde biomaterial fixation.[8]

## 2. Materials and methods

### 2.1. In-house isolation of type I collagen from rat tail tendons

Type I collagen was isolated in-house via acidic treatment of rat tail tendons.[29] Briefly, frozen rat tails were thawed in ethanol (for about 20 min). Individual tendons were pulled out of the tendon sheath, minced, and placed in 17.4 mM acetic acid (Sigma Aldrich) solution at 4 °C in order to extract collagen. After three days extraction, the mixture was centrifuged at 20000 rpm for one hour. The pellet was discarded and the crude collagen solution was neutralised with

0.1 M NaOH (Sigma Aldrich). After stirring (overnight, 4 °C), the neutralised solution was centrifuged (45 min, 10000 rpm, 4 °C). The supernatant volume was measured and an equal volume of fresh acetic acid (17.4 mM) solution was used to re-solubilise the collagen pellet. The mixture was then freeze-dried in order to obtain collagen. The resulting product was analysed via sodium dodecyl sulphate-polyacrylamide gel electrophoresis (SDS-page) which showed only the main electrophoretic bands of type I collagen.[30]

## 2.2. Synthesis of collagen hydrogels

In-house isolated type I collagen (0.8 wt.-%) was dissolved in 10 mM hydrochloric acid (Sigma Aldrich) under stirring at room temperature, prior to network formation via either 1,3-Phenylenediacetic acid (Ph, VWR International) or N-(3-Dimethylaminopropyl)-N′-ethylcarbodiimide hydrochloride (EDC, Sigma Aldrich). At the same time, Ph was stirred (0°C, 30 min) in sodium phosphate buffer (0.1 M, pH 7.4, 500 µL) at selected molar ratios to target collagen lysines (0.5-1.5 COOH/Lys molar ratio).[31] Here, a three-fold molar content of EDC and N-Hydroxysuccinimide (NHS) was added. NHS-activated Ph solution was mixed with obtained collagen solution (1 g), so that the pH of the resulting mixture was 6.5 at 19.2 °C (in the case of 1 COOH/Lys molar ratio). Reacting mixtures were incubated overnight under gentle shaking at room temperature, in order to allow for the nucleophilic addition reaction of collagen lysines to Ph carboxylic functions to occur. EDC-crosslinked collagen was synthesised as state-of-the-art crosslinked collagen control by mixing EDC with obtained collagen solution, as previously reported.[19] Complete gel formation was observed following reaction with either Ph or EDC. Resulting hydrogels were washed with distilled water and dehydrated in aqueous solutions of increasing ethanol concentrations.

## 2.3. Investigation of chemical composition and structural organisation

Attenuated Total Reflectance Fourier-Transform Infrared (ATR FT-IR) was carried out on dry samples using a Perkin-Elmer Spectrum BX spotlight spectrophotometer with diamond ATR attachment. Scans were conducted from 4000 to 600 $cm^{-1}$ with 64 repetitions averaged for each spectrum. Resolution was 4 $cm^{-1}$ and interval scanning was 2 $cm^{-1}$.

## 2.4. 2,4,6-Trinitrobenzenesulfonic acid (TNBS) assay

The degree of collagen crosslinking was determined on dry collagen networks via 2,4,6-trinitrobenzenesulfonic acid (TNBS) colorimetric assay.[32] 11 mg of dry sample were mixed with 1 mL of 4 wt.-% $NaHCO_3$ (pH 8.5) and 1 mL of 0.5 wt.-% TNBS solution at 40 °C under mild shaking. After 4 hours reaction, 3 mL of 6 M HCl solution were added and the mixture was heated to 60 °C to dissolve any sample residuals. Solutions were cooled down and extracted three times with anhydrous ethyl ether to remove non-reacted TNBS species. All samples were read against a blank, prepared by the above procedure, except that the HCl solution was added before the addition of TNBS. The content of free amino groups and degree of crosslinking (*C*) were calculated as follows:

$$\frac{moles(Lys)}{g(collagen)} = \frac{2 \times Abs(346\ nm) \times 0.02}{1.4 \times 10^4 \times b \times x} \quad \textbf{(Equation 1)}$$

$$C = \left(1 - \frac{moles(Lys)_{Crosslinked}}{moles(Lys)_{Collagen}}\right) \times 100 \quad \textbf{(Equation 2)}$$

where *Abs(346 nm)* is the absorbance value at 346 nm, $1.4 \cdot 10^4$ is the molar absorption coefficient for 2,4,6-trinitrophenyl lysine (in L/mol·$cm^{-1}$), *b* is the cell path length (1 cm), *x* is the sample weight, and *moles(Lys)$_{Crosslinked}$* and *moles(Lys)$_{Collagen}$* represent the lysine molar content in crosslinked and native collagens. Two replicas were used for each sample composition, whereby *C* results were described as average ± standard deviation.

## 2.5. Wide Angle X-ray Scattering

Wide Angle X-ray Scattering (WAXS) was carried out on dry collagen networks with a Bruker D8 Discover (40 kV, 30 mA, x-ray wavelength $\lambda$ = 0.154 nm). The detector was set at a distance of 150 mm covering *2θ* from 5 to 40°. The collimator was 2.0 mm wide and the exposure time was 10 s per frame. Collected curves were subtracted from the background (no sample loaded) curve and fitted with polynomial functions.

## 2.6. Swelling tests

Swelling tests were carried out by incubating dry samples in 5 mL distilled water for 24 hours. Water-equilibrated samples were retrieved, paper-blotted and weighed. The weight-based swelling ratio (*SR*) was calculated as follows:

$$SR = \frac{m_s - m_d}{m_d} \times 100 \qquad \textbf{(Equation 3)}$$

where $m_s$ and $m_d$ are swollen and dry sample weights, respectively. Three replicas were used for each sample composition, so that *SR* results were expressed as average ± standard deviation.

## 2.7. Thermal analysis

Differential Scanning Calorimetry (DSC) temperature scans were conducted on formed collagen hydrogels in the range of 10-140 °C with either 10 or 1°C·min$^{-1}$ heating rate (TA Instruments Thermal Analysis 2000 System and 910 Differential Scanning Calorimeter cell base). The DSC cell was calibrated using indium with 20 °C·min$^{-1}$ heating rate under 50 cm$^3$·min$^{-1}$ nitrogen atmosphere. 5-10 mg sample was applied in each measurement.

**2.8. Compression tests**

Water-equilibrated hydrogel discs (ø 0.8 cm) were compressed at room temperature with a compression rate of 3 mm·min$^{-1}$ (Instron 5544 UTM). A 500 N load cell was operated up to sample break. The maximal compressive stress ($\sigma_{max}$) and compression at break ($\varepsilon_b$) were recorded, so that the compressive modulus ($E$) was calculated by fitting the linear region of the stress-strain curve. Four replicas were employed for each composition and results expressed as average ± standard deviation.

**2.9. Hydrolytic degradation**

The hydrolytic degradation of crosslinked samples was investigated via 1-week incubation in PBS (pH 7.4, 25 °C) (Lonza). Retrieved samples were rinsed with distilled water, dried and weighed. The relative mass change ($M_R$) in retrieved samples was determined as:

$$M_R = \frac{m_t - m_d}{m_d} \times 100 \qquad \textbf{(Equation 4)}$$

whereby $m_t$ and $m_d$ correspond to the dry sample mass after and before PBS incubation, respectively. Two replicas were used for each composition so that $M_R$ was described as average ± standard deviation. Another sample replica was used for SEM analysis (JEOL SM-35), whereby any micro-structural alteration was investigated on gold-coated, recovered samples.

## 2.10. Bioactivity study

A mineralisation experiment was carried out via a 1-week sample incubation at 25 °C in SBF, with a sample weight to SBF volume ratio of 0.07 g 50 ml$^{-1}$.[33] Retrieved samples were rinsed with distilled water, dried, weighed and analysed via ATR-FTIR. The relative mass change ($M_R$) in retrieved samples was determined via Equation 4. Four replicas were used for each composition so that $M_R$ was described as average ± standard deviation. Following gold-coating, sample morphological investigations were carried out via SEM and EDS (JEOL SM-35), in order to identify any structural change in recovered samples and the chemical composition of potential mineral phase deposited on the material.

An additional incubation test in which samples were incubated in a calcium chloride solution (2.5 mM $CaCl_2$, pH 7.4, 25 °C) for two days was carried out in order to explore the mechanism of calcium phosphate deposition on Ph-crosslinked collagen hydrogels, aiming to elucidate any interaction between solution calcium species and the Ph aromatic ring. Retrieved samples were rinsed in distilled water and dried before ATR-FTIR analysis.

## 2.11. Extract cytotoxicity study

A cytotoxicity assay was conducted on freshly synthesised ethanol-treated collagen hydrogels following European norm standards (EN DIN ISO standard 10993-5). 0.1 mg of hydrogel sample was incubated in 1 mL cell culture medium (Dulbecco's Modified Eagle Medium, DMEM) at 37 °C. After 72-hour incubation, sample extract was recovered by centrifugation and applied to 80% confluent L929 mouse fibroblasts cultured on a polystyrene 96-well plate. Dimethyl sulfoxide (DMSO) was used as negative control, while cell culture medium was used as positive control. Cell morphology was monitored using a transmitted light microscope (phase contrast mode, Zeiss, Germany) and coupled with a colorimetric cytotoxicity assay (CellTiter 96® AQ$_{ueous}$ Assay, Promega), in order to quantify the number of viable cells in proliferation. In the presence of phenazine methosulfate (PMS), 3-(4,5-dimethylthiazol-2-yl)-5-(3-carboxymethoxyphenyl)-2-(4-sulfophenyl)-2H-tetrazolium (MTS) is bioreduced by cells into a formazan product that is soluble in tissue culture medium. The conversion of MTS into aqueous, soluble formazan is accomplished by dehydrogenase enzymes found in metabolically active cells. Thus, the quantity of formazan product as measured by the amount of 490 nm absorbance is directly proportional to the number of living cells in culture.

## 3. Results and discussion

Collagen hydrogels were prepared via lysine functionalisation with Ph as stiff, aromatic segment (Scheme 1), aiming at establishing a reliable, alternative synthetic strategy for the design of collagen-based, bone tissue-like scaffolds. Following nucleophilic addition of lysine terminations to NHS-activated carboxylic functions of Ph, a covalent network consisting of newly-formed, hydrolytically-cleavable, amide bond net-points, was expected. Complete gel formation was observed following incubation of collagen solution with Ph-containing mixture. Freshly-synthesised hydrogels were thoroughly washed with distilled water in order to remove any non-reacted moiety and dehydrated with increasing ethanol solution series for subsequent characterisation. In the following, molecular architecture, macroscopic properties, and material functionalities in obtained hydrogels will be described and compared to the state-of-the-art EDC-crosslinked collagen materials. Sample are coded as *Collagen-XXX-YY*, where *XXX* indicates the type of system (either Ph or EDC-based), while *YY* identifies the molar ratio of either Ph carboxylic functions or EDC to collagen lysines.

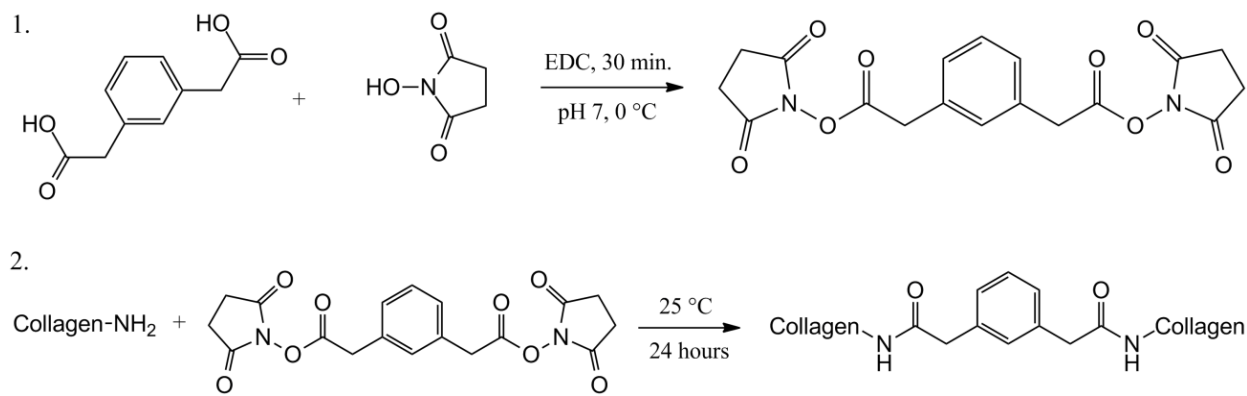

**Scheme 1.** Formation of collagen hydrogels via lysine functionalisation with a bifunctional aromatic segment. Ph is NHS-activated in presence of EDC (1.) prior to reaction with collagen (2.). Lysine functionalisation occurs via nucleophilic addition to activated carboxylic functions, resulting in the formation of a covalent network.

### 3.1. Chemical and structural analysis in covalent networks

ATR-FTIR spectral analysis was carried out in order to explore the chemical composition and molecular conformation in resulting collagen networks. Collagen displays distinct amide bands via FTIR, which can provide information on its triple helix structure.[25,34,35] These are: (i) amide A and B bands at 3300 and 3087 cm$^{-1}$, respectively, which are mainly associated with the stretching vibrations of N-H groups; (ii) amide I and II bands, at 1650 and 1550 cm$^{-1}$, resulting from the stretching vibrations of peptide C=O groups as well as from N–H bending and C–N stretching vibrations, respectively; (iii) amide III band centered at 1240 cm$^{-1}$, assigned to the C–N stretching and N–H bending vibrations from amide linkages, as well as wagging vibrations of CH$_2$ groups in the glycine backbone and proline side chains. Each of the previously-mentioned amide bands are exhibited in FTIR spectra of both Ph- and EDC-crosslinked samples, whereby no detectable band shift is displayed compared to the spectrum of native type I collagen (Figure 1). This observation suggests that the collagen triple helical organisation can be preserved in formed covalent networks. Other than qualitative findings on unchanged band positions, FTIR absorption ratio of amide III to 1450 cm$^{-1}$ band ($A_{III}/A_{1450}$) was determined in both crosslinked and native collagens, as well as in gelatin as a denatured collagen control sample, in order to quantify the degree of triple helix preservation following network formation. As in the case of native type I collagen, all systems proved to show an amide ratio close to unity ($A_{III}/A_{1450}$ = 1.01-1.14), while a lowered amide ratio ($A_{III}/A_{1450}$ = 0.84) was found in the case of the gelatin sample. These results provide evidence of the preserved triple helix integrity[34] following functionalisation of native collagen, which was not observed in the case of gelatin, as a representative sample of denatured collagen.

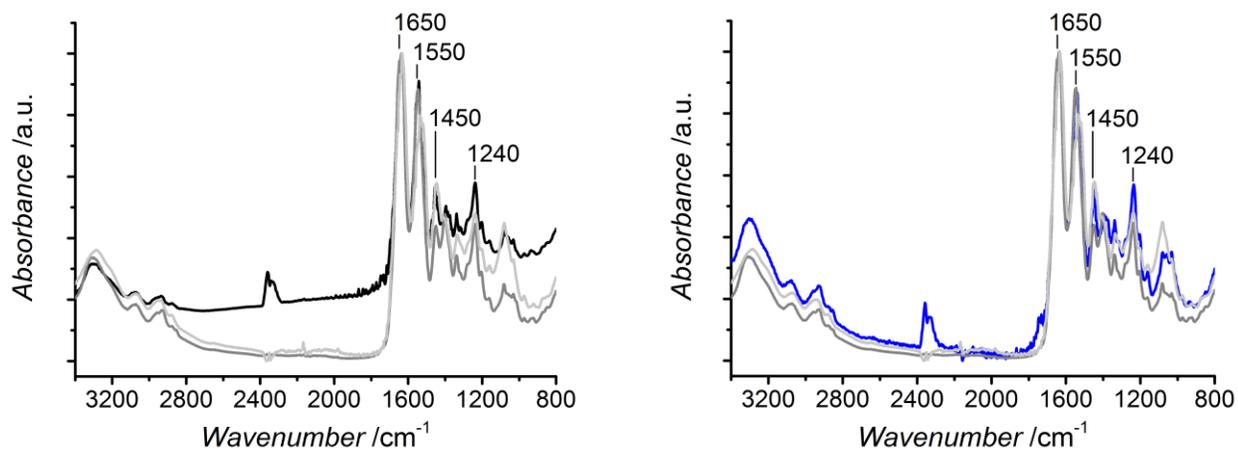

**Figure 1.** Exemplary ATR-FTIR spectra of Ph- (—, left) and EDC- (—, right) crosslinked collagens. Native collagen (—) and gelatin (—) spectra are displayed as controls in both plots.

Given that the triple helix absorption ratio was determined based on the intensity of amide bands and given that these bands are present in all protein regardless of their conformation state, WAXS was applied in order to further confirm the protein conformation in resulting networks. WAXS has been previously applied in order to elucidate information about the packing features of collagen in terms of distances between collagen molecules in the lateral plane of the collagen fibril (intermolecular lateral packing) and distances between amino acids along the polypeptide chain (helical rise per residue).[35-37] Figure 2 displays WAXS spectra of linear intensity vs. scattering vector resulting from type I collagen and samples Collagen-Ph1 and Collagen-EDC60. As expected, WAXS spectrum of type I collagen displays three main collagen peaks, identifying the intermolecular lateral packing of collagen molecules ($d \sim 1.1$ nm, $2\Theta \sim 8°$), the isotropic amorphous region ($d \sim 0.5$ nm, $2\Theta \sim 20°$) and the axial periodicity ($d \sim 0.29$ nm, $2\Theta \sim 31°$) of polypeptide subunits *(Gly-X-Y)* along a single collagen chain. Similarly, the sample Collagen-Ph1 reveals a similar WAXS spectrum, so that the 1.1 nm peak corresponding to the triple helix packing is still present following network formation. In contrast to that, the spectrum of EDC-crosslinked sample shows detectable alterations compared to the one of native collagen. Here, the peak related to intermolecular lateral packing of collagen molecules is less pronounced.

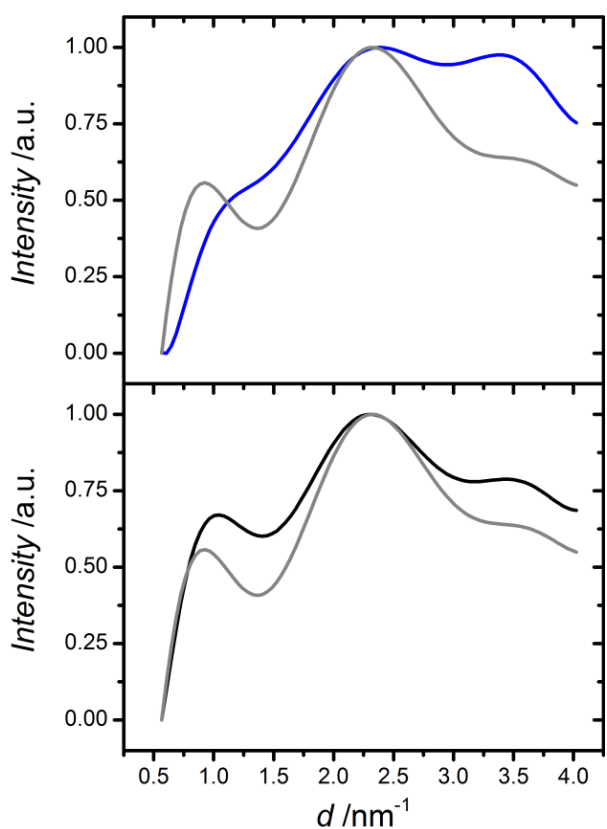

**Figure 2.** WAXS spectra of samples Collagen-Ph1 (—), Collagen EDC60 (—) and in-house isolated type I collagen (—).

These WAXS observations are not in agreement with previous ATR-FTIR results, suggesting an alteration in the native collagen packing in EDC-crosslinked samples. At the same time, a preserved collagen conformation is confirmed in Collagen-Ph samples. This is an interesting result, since aromatic residues are expected to partially destabilise the collagen trimers due to their inability to form hydrogen bonds, as recently observed in the case of 4-vinyl benzyl chloride (4VBC)-functionalised type I collagen.[30] However, amide bonds are formed following reaction of collagen lysines with NHS-activated Ph, potentially mediating hydrogen bonds among collagen molecules. Consequently, despite the presence of destabilising Ph aromatic moieties, triple helix structure can still be preserved in resulting collagen networks.

### 3.2. Network architecture in triple-helical collagen hydrogels

Following elucidation of protein conformation, network architecture was investigated in resulting materials by both quantifying the degree of crosslinking (*C*) via TNBS assay and by assessing the swelling ratio (*SR*). Here, it was of interest to understand whether network architecture, i.e. crosslinking density, could be adjusted via systematic variation of molecular parameters, i.e. Ph/Lys molar ratio, so that hydrogel properties could be controlled.

As observed in Table 1, collagen-Ph samples displayed a higher degree of crosslinking (87-99 mol.-%) compared to state-of-the-art EDC-crosslinked collagen (25-68 mol.-%), despite the fact that a much lower Ph-Lys, compared to EDC/Lys, molar ratio was applied. Here, a slight increase of Ph content (0.5 to 1.5 [COOH]/[Lys] ratio) in the crosslinking mixture led to nearly-complete functionalisation of collagen lysines (> 99 mol.-%). In order to explain the different yield of crosslinking density in Ph- and EDC-based systems, the network architecture in both systems must be carefully considered. In Ph-crosslinked collagen, the employment of Ph as bifunctional segment is likely to promote crosslinking of collagen molecules separated by a distance.[22] On the other hand, EDC-mediated functionalisation results in the formation of zero-length net-points, whereby only intramolecular crosslinks are likely established.[19,20]

Consequently, steric hindrance effects are likely to explain the decreased degree of crosslinking observed in collagen-EDC, while an increased functionalisation is promoted in collagen-Ph, samples.

**Table 1.** Degree of crosslinking (*C*), swelling ratio (*SR*) and denaturation temperature ($T_d$) in collagen hydrogels, as determined via TNBS assay, swelling tests, and DSC analysis, respectively.

| Sample ID | *C* /mol.-% | *SR* /wt.-% | $T_d$ /°C |
|---|---|---|---|
| Collagen-Ph0.5 | 88 ± 3 | 1285 ± 450 | 80 |
| Collagen-Ph1 | 87 ± 14 | 1311 ± 757 | 80 |
| Collagen-Ph1.5 | > 99 | 823 ± 140 | 88 |
| Collagen-EDC10 | 25 ± 6 | 1392 ± 82 | 68 |
| Collagen-EDC20 | 37 ± 13 | 1595 ± 374 | 76 |
| Collagen-EDC30 | 34 ± 1 | 1373 ± 81 | 78 |
| Collagen-EDC40 | 68 ± 3 | 1374 ± 182 | 80 |
| Collagen-EDC60 | 60 ± 1 | 1106 ± 82 | 80 |

In order to further confirm TNBS findings, hydrogel swelling ratio was also investigated. Comparing the whole set of sample compositions, the swelling ratio of sample collagen-Ph1.5 was found to be significantly lower compared to all other EDC-based samples (Table 1). This is in agreement with TNBS results, suggesting nearly complete functionalisation of collagen lysines for this composition. As for the other Ph-crosslinked samples (collagen-Ph0.5/1), swelling ratios were similar to each other, which is again supported by TNBS results, describing a similar degree of crosslinking. Either *C* or *SR* results therefore confirmed that network architecture and crosslinking density was successfully adjusted in Ph-based collagen systems, whereby collagen molecules were covalently functionalised with preserved triple helical conformation (Figure 3).

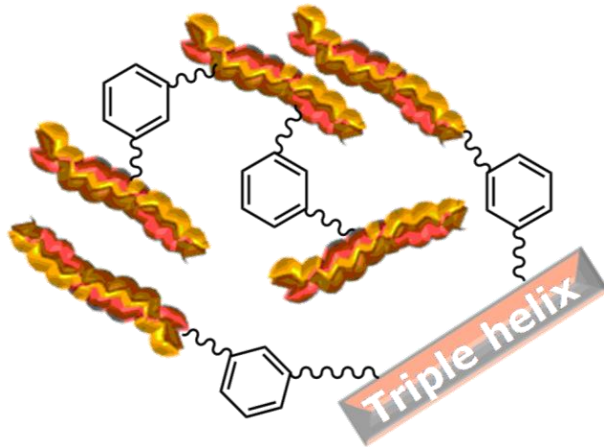

**Figure 3.** Network architecture in triple-helical collagen systems. Collagen is Ph-functionalised in diluted acidic conditions, resulting in intermolecular crosslinking of collagen triple helices. Formed hydrogels highlight preserved triple helical organisation, while macroscopic properties can be controlled by variation in network architecture.

### 3.3. Thermo-mechanical properties of collagen hydrogels

Thermo-mechanical analysis was carried out on collagen hydrogels in order to investigate whether variation of molecular parameters could induce changes in macroscopic properties. Thus, DSC was performed aiming at investigating the thermal stability of formed hydrogels in physiological conditions and combined with compression tests in order to study the material mechanical behaviour in aqueous environment.

Thermal stability of collagen is dictated by its denaturation temperature ($T_d$), which is related to the unfolding of collagen triple helices into randomly-coiled chains; $T_d$ is therefore expected to be highly affected by the presence of covalent net-points among collagen molecules.[9,22,25] Figure 4 (left) depicts exemplary thermograms of native, Ph- and EDC-crosslinked type I collagens, all describing an endothermic thermal transition in the range of 67-88 °C. In the case of native type I collagen, such endothermic transition was recorded at around 67 °C, which is in agreement with previous reports describing a thermal denaturation of collagen at 55-67 °C (using similar heating rates),[25,30] although decreased values (at ca. 34 °C) were also observed at lowered heating rates (i.e. 0.1-2 °C·min$^{-1}$ heating rate).[35] Considering the influence of DSC heating rate on the peak position of collagen denaturation temperature (50→100 °C at 0.5→5 °C·min$^{-1}$),[36] the endothermic transitions recorded in this study are likely to identify the thermal denaturation of collagen. In order to elucidate this point further, ATR-FTIR was recorded on a DSC-treated sample, aiming to confirm that the observed thermal transition was not related to a decomposition process. As observed in Figure 4 (right), ATR-FTIR spectrum still shows the main collagen peaks following DSC analysis, giving supporting evidence that no decomposition process occurred in the sample. These investigations thus confirmed that the obtained DSC thermograms effectively displayed an endothermic transition ascribed to collagen denaturation. These findings provided further evidence of the preserved integrity of triple helices in formed hydrogels, as already pointed out by ATR-FTIR (Figure 1) and WAXS (Figure 2). Remarkably, the collagen denaturation temperature is shifted towards higher temperature in crosslinked compared (68-88 °C) to native collagen (67 °C) samples (Table 1), whereby a slight influence of DSC heating rate on $T_d$ was also observed (Figure 4, left). Considering previous composition-dependent changes in network architecture, variation of $T_d$ seems to be directly related to changes of crosslinking degree in the hydrogel networks. These results give supporting evidence that covalent net-points were established during hydrogel formation, so that collagen triple helices were successfully retained and stabilised. Remarkably, Ph-crosslinked collagen proved to display higher denaturation temperatures with respect to EDC- ($T_d$: 82 °C, $C$: 50 mol.-%),[25] glutaraldehyde- ($T_d$: 78 °C, $C$: 51 mol.-%),[21] and hexamethylene diisocyanate- ($T_d$: 74 °C, $C$: 48 mol.-%)[23] crosslinked collagen materials. This indicates that the incorporation of Ph as a stiff, aromatic segment superiorly stabilises collagen molecules in comparison with current crosslinking methods.

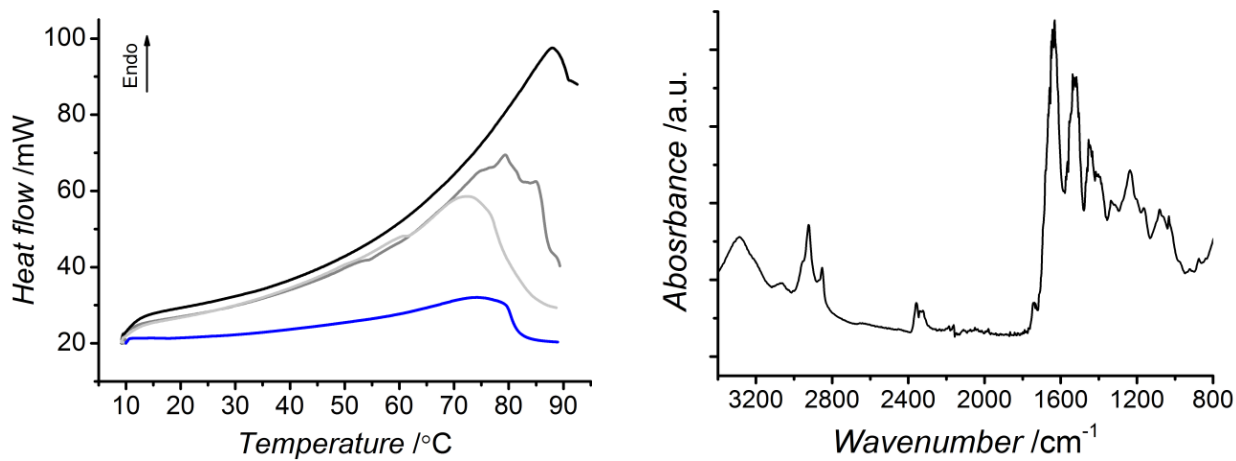

**Figure 4.** Left: DSC thermograms of samples Collagen-Ph1.5 (—), Collagen-EDC60 (—) and type I collagen (—) obtained with 10 °C·min$^{-1}$ heating rate. An additional thermogram of sample Collagen-Ph1 was acquired with 1°C·min$^{-1}$ heating rate (—). All samples describe a triple helix denaturation by an endothermic transition at 67-88 °C. Right: ATR-FTIR spectrum of sample Collagen-Ph1 following DSC analysis; the main peaks of collagen can still be observed in the spectrum.

Besides thermal analysis, mechanical properties of collagen-Ph hydrogels were measured by compression tests. Samples described *J*-shaped stress-compression curves (Figure 5), similar to the case of native tissues.[9] Here, shape recovery was observed following load removal up to nearly 50% compression, suggesting that the established covalent network successfully resulted in the formation of an elastic material, as observed in linear biopolymer networks.[37] On the other hand, EDC-crosslinked collagen showed minimal mechanical properties, whereby sample failure was observed even after sample punching. Consequently, quantitative data of mechanical properties on collagen-EDC samples could not be acquired. These findings support the idea that the collagen functionalisation with Ph is effective for the formation of collagen materials with enhanced mechanical properties.

Compressive modulus ($E$: 28±10→35±9 kPa) and maximal stress of Ph-crosslinked samples ($\sigma_{max}$: 6±2→8±4 kPa) were measured in the kPa range, while compression at break ($\varepsilon_b$: 53±5→58±5 %) did not exceed 60% compression (Figure 6). The resulting compressive modulus was therefore measured to be almost 20 times higher compared to previously-reported collagen-based materials.[22]

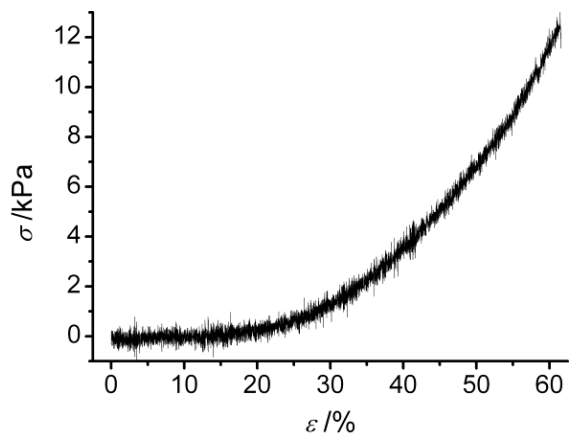

**Figure 5.** Compressive stress-strain plot of hydrogel Collagen-Ph1 compressed up to sample break describing a *J*-shaped curve, typical of native tissues. Shape recovery was observed following compression up to nearly 50%.

At the same time, there was little variation in mechanical properties among the different compositions. Given the unique organisation of collagen, the hierarchical level, e.g. triple helix, fibrils or fibres, at which covalent crosslinks are introduced, is crucial in order to study the influence of crosslinking on the mechanical properties of collagen. Olde Damink et al. observed no variation of mechanical properties in dermal sheep crosslinked collagen.[20,21,23] This was explained based on the fact that crosslinks were mainly introduced within rather than among collagen molecules. This hypothesis may be supported by above mechanical findings, although it is not in line with TNBS, swelling and thermal analysis data. Most likely, the variation of Ph feed ratio among the different compositions (0.5→1.5 [COOH]/[Lys] ratio) was probably too low to result in significant changes in mechanical properties. For these reasons, a wider range of Ph concentrations may be advantageous in order to establish hydrogels with varied mechanical properties. In that case, investigation via AFM will be instrumental in order to explore the hierarchical levels (i.e. triple helices, fibrils or fibres) at which covalent crosslinks are introduced; here, tapping mode[12] could be employed in order to scan the material surface and explore whether any collagen structural assembly is present following covalent crosslinking. Subsequently, force-volume measurements could give information on the local elasticity of resulting crosslinked patterns.

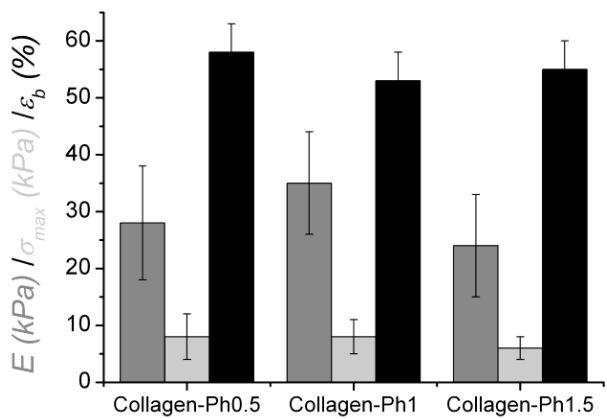

**Figure 6.** Compressive modulus ($E$), maximal compressive stress ($\sigma_{max}$), and compression at break ($\varepsilon_b$) of Ph-crosslinked collagen hydrogels.

### 3.4. Degradability and bioactivity *in vitro*

SBF incubation is a well-known method to test a material's ability to form a hydroxycarbonate apatite (HCA) layer *in vitro*.[33] Collagen is known to trigger bone-like apatite deposition *in vivo* during bone formation,[11] so it was of interest to investigate collagen hydrogel behaviour in SBF as an osteogenic-like medium. The mass change was therefore quantified and the presence of calcium/phosphorous elements in retrieved samples assessed in order to (i) clarify any occurrence of degradation and mineral deposition and (ii) determine the chemical composition of any potentially nucleated phases. This investigation was then coupled to a degradation study in PBS in order to investigate the stability of collagen hydrogels in a calcium-free medium.

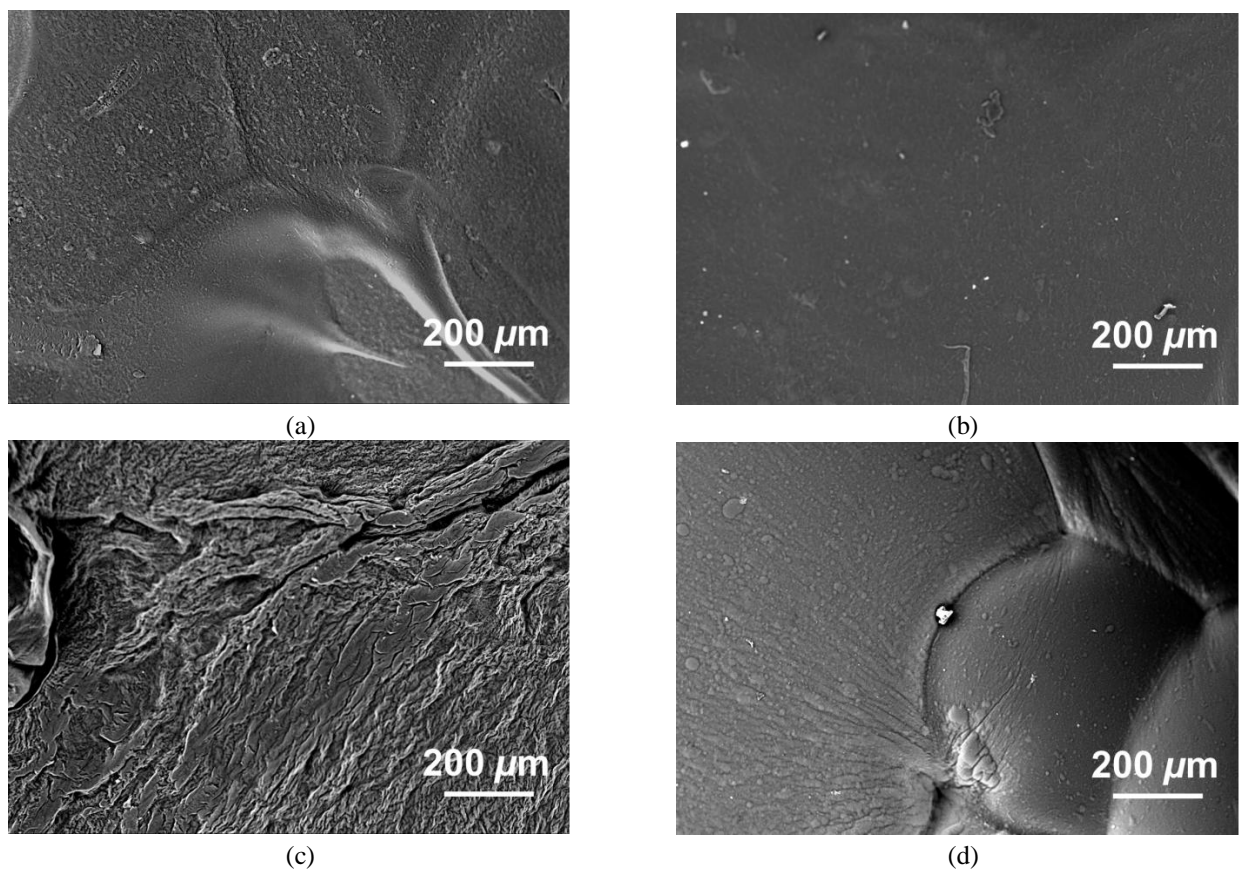

**Figure 7.** SEM on samples Collagen-Ph1 (a), Collagen-Ph1.5 (b), Collagen-EDC10 (c) and Collagen-EDC60 (d), following 1-week incubation in PBS (pH 7.4, 25 °C). Retrieved EDC-crosslinked samples present damaged surface, providing supporting evidence of hydrolytic degradation taking place in these materials. The extent of surface damage seems to be directly related to the degree of crosslinking. In contrast to that, Ph-crosslinked samples reveal no detectable damage on their surfaces.

A slight mass decrease (averaged mass loss ~ 8 wt.-%) was observed in PBS-retrieved collagen-Ph samples, indicating minimal hydrolytic degradation had occurred (Table 2), in contrast to EDC-crosslinked samples ($M_R$ > 30 wt.-%). The latter high change in mass was also reflected by SEM of partially-degraded EDC controls (Figure 7), whereby a crack-like formation was observed following PBS incubation. In contrast to that, the collagen-Ph1.5 sample described the lowest mass loss and a nearly-intact material microstructure, confirming that increased Ph/Lys feed molar ratio led to networks with increased crosslinking degree (Table 1). Interestingly, no specific fibrillar pattern was observed in resulting collagen samples. This likely suggests that Ph-mediated covalent crosslinking hinders reconstitution of collagen fibres, so that a triple helical architecture is expected in formed networks (Figure 3), as evidenced by ATR-FTIR (Figure 1), WAXS (Figure 2) and DSC (Figure 4) analyses.

**Table 2.** Mass change ($M_R$) in retrieved samples following 1-week incubation in either PBS or SBF. The negative mass change describes a mass loss in recovered samples likely due to hydrolytic degradation of the material in the tested conditions. EDS *Ca/P* ratio is presented for samples treated in SBF.

| *Sample ID* | $M_R$ /wt.-% | | *Ca/P* |
|---|---|---|---|
| | PBS | SBF | |
| Collagen-Ph0.5 | − (10 ± 8) | − (5 ± 9) | 1 |
| Collagen-Ph1 | − (10 ± 5) | 5 ± 5 | 0.84 |
| Collagen-Ph1.5 | − (5 ± 8) | − (11 ± 5) | 1.41 |
| Collagen-EDC10 | − (38 ± 30) | − (48 ± 13) | 1.26 |
| Collagen-EDC60 | − (33 ± 4) | − (58 ± 19) | 1.09 |

Besides PBS, gravimetric analysis in SBF-treated samples confirmed similar trends, although one sample, collagen-Ph1, displayed a slight mass increase, suggesting deposition of a nucleated phase on the material. SEM on retrieved Ph-based samples revealed nearly-intact material surfaces, confirming that a covalent network was still present at the molecular level (Figure 8). Other than Ph-based systems, a much higher mass loss (averaged mass loss ~ 53 wt.-%) was observed in EDC-crosslinked collagen, supporting *C* and *SR* findings. Furthermore, small (up to 1 $\mu$m) micro-pores were observed on the retrieved sample surface, which is likely to be related to the material degradation, similarly to the case of PBS-retrieved samples (Figure 8).

Following gravimetric and morphological investigations, FTIR and EDS were carried out on SBF-retrieved samples in order to explore the chemical composition of retrieved samples. Figure 9 describes the spectra of the collagen-Ph1.5 sample before and after SBF incubation.

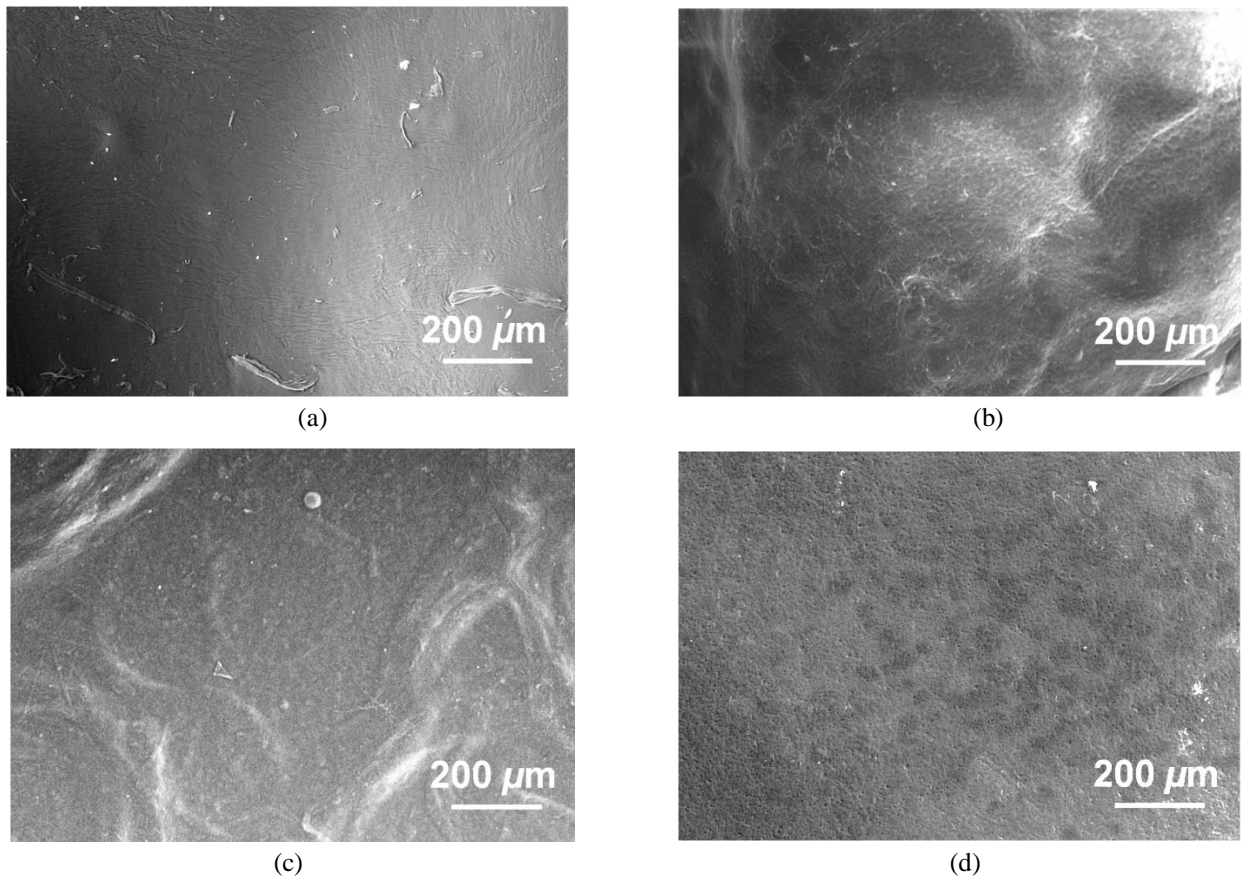

**Figure 8.** SEM on samples collagen-Ph1 and collagen-EDC10 prior to (a, b) and following (c, d) 1-week incubation in 25 °C SBF, respectively. Minimal surface damage is observed in retrieved Ph-crosslinked samples (c), while the formation of pores is displayed in resulting EDC-crosslinked samples (d), likely ascribed to the hydrolytic degradation of these materials.

Main amide bands are detected in both collagen spectra, proving that collagen composition was not affected following *in vitro* bioactivity test. Here, additional bands at 1078, 1030 and 970 cm$^{-1}$ were observed in spectra of SBF-incubated hydrogels. These bands are characteristic of phosphate species associated with apatite.[38] At the same time, a slight shift is described by the Amide II band in FTIR spectrum of SBF-incubated sample in comparison to freshly-synthesised hydrogel. Given that the Amide II band is located in the same region as the C=C band (1540 cm$^{-1}$) of Ph aromatic ring, this shift may suggest a specific interaction between the introduced Ph segment and calcium species of nucleated phase.[39] This hypothesis is also suggested by the shape of this peak, which is broader compared to the same peak in a FTIR spectrum of a freshly-synthesised sample. In order to further explore the role of the Ph aromatic ring in templating calcium nucleation, the collagen-Ph1.5 sample was incubated in a simple calcium containing solution (2.5 mM $CaCl_2$). Interestingly, in the resulting ATR-FTIR spectrum on the recovered sample (Figure 9), the 1540 $cm^{-1}$ peak was absent, in contrast to native type I collagen spectrum. This observation may give further evidence of an interaction between calcium species present in solution and the aromatic Ph ring present in the collagen network. Although further studies are needed to confirm this point, these results support the hypothesis that introduced aromatic moieties may act as a preferential site of calcium phosphate nucleation.

Complementing FTIR, EDS analysis was carried out in order to further characterise the chemical composition of hydrogel nucleated phase following SBF incubation. In doing this, complete sample washing with water was crucial in order to remove superficial deposition of magnesium, sodium and chlorine ions. The presence of calcium and phosphorous elements was observed in all samples, although with a low *Ca/P* atomic ratio (*Ca/P*: 0.84-1.41, Table 2). This suggests that the mineral phase laid down on the material surface was most likely constituted of amorphous rather than crystalline calcium phosphate, which may be expected due to the relatively short incubation (1 week) at 25 °C. The amorphous nature of the hydrogel nucleated phase is also supported by the broad phosphate peaks in FTIR spectra of resulting samples (Figure 9).

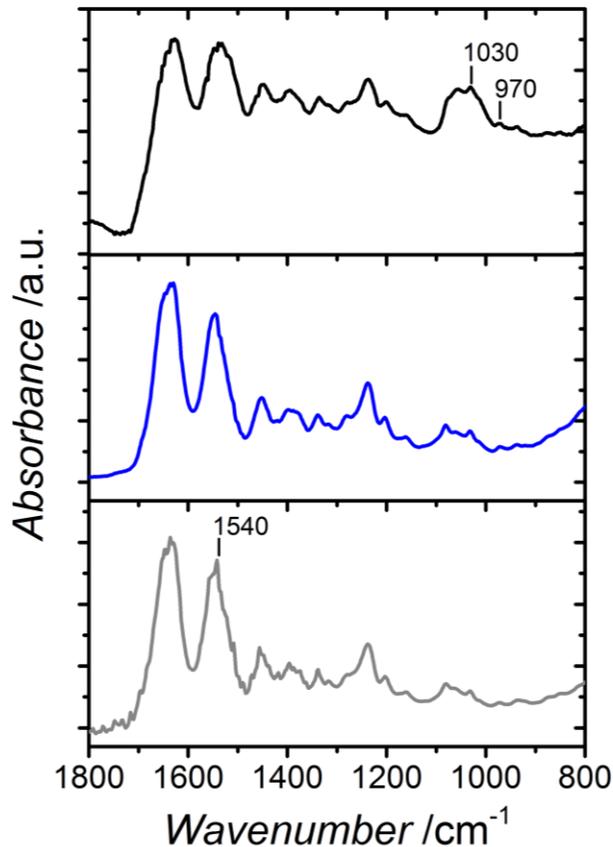

**Figure 9.** Exemplary ATR-FTIR spectra of sample Collagen-Ph1.5 directly after synthesis (—), following 2-day incubation in 2.5 mM $CaCl_2$ solution (—), and after 1-week incubation in SBF medium (—). SBF-retrieved samples display additional peaks at 970 and 1030 $cm^{-1}$, distinctive of phosphate phase deposition, while disappearance of the peak related to aromatic bands (at 1540 $cm^{-1}$) is observed after incubation in either $CaCl_2$ or SBF solutions.

The collagen-Ph1.5 sample displayed the highest *Ca/P* atomic ratio (*Ca/P* ~ 1.41) compared to all other samples. This finding may provide further evidence of a selective mechanism of apatite nucleation in hydrogels crosslinked with increased Ph content, as suggested by FTIR (Figure 9). Here, incorporated aromatic rings are expected to act as selective sites of calcium-phosphate nucleation following chelation with calcium ions from SBF, similarly to the interaction mechanism observed in cortical bone between calcium ions and glycosaminoglycan sulfate and carboxylate groups.[40] Although further investigations with longer time at body temperature are required in order to elucidate the role of Ph in guiding matrix-nucleated phase interaction, this study provides experimental evidence that these systems could promote calcium-phosphate nucleation *in vitro* and therefore may be suitable as implants for the regeneration of mineralised tissues.

### 3.5. Cytotoxicity study

Material cyto-compatibility was investigated by extract cytotoxicity assay following European guidelines for medical device testing. Samples synthesized with the lowest and highest Ph feed ratio were tested, in order to investigate whether any toxic compound could be released from formed hydrogels to cell culture medium following sample incubation. Cell morphology was monitored at 24 and 48 hours following cell culture in both sample compositions as well as in positive (DMEM) and negative controls (DMSO); these observations were then complemented with MTS assay quantitative data of cell viability after 48 hours. As pictured in Figure 10, L929 mouse fibroblasts appeared confluent when cultured in either sample extracts or cell culture medium, suggesting that sample extracts were well tolerated by cells. On the other hand, cell death was observed in the case of cell culture in DMSO. These morphology observations were in agreement with MTS assay results, describing the metabolic activity of vital cultured cells (Figure 11). Interestingly, measured formazan content was higher in the case of cells cultured in both samples extracts compared to the case of cell culture in DMEM and DMSO.

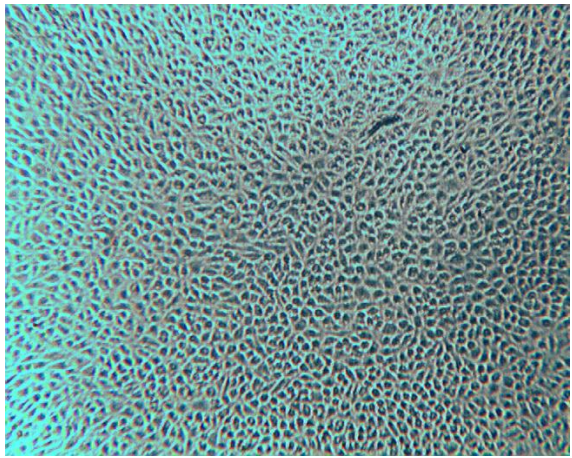
(a)

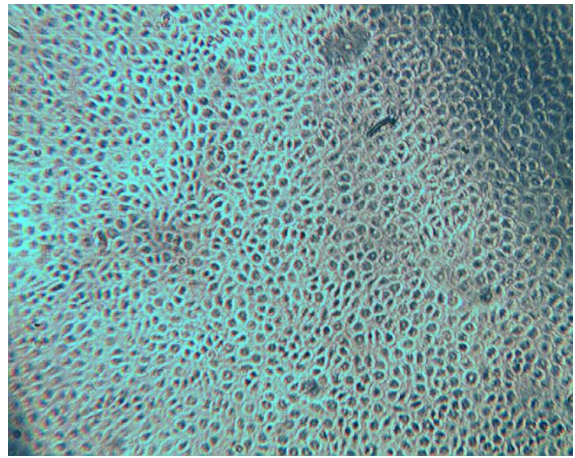
(b)

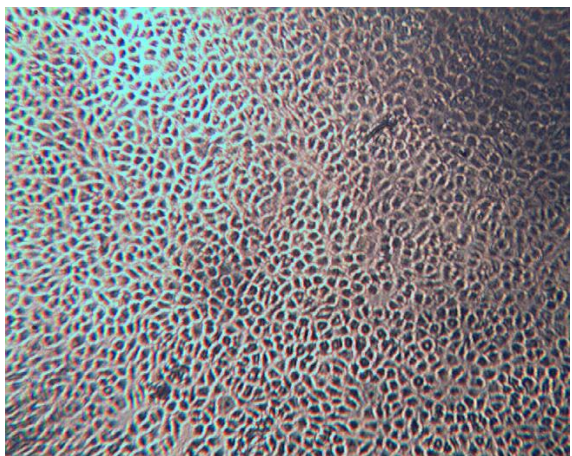
(c)

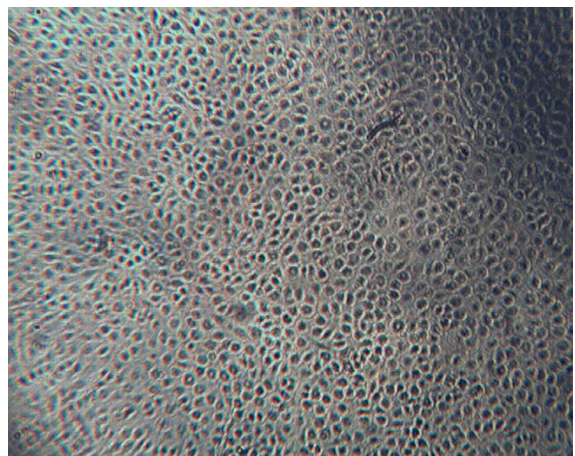
(d)

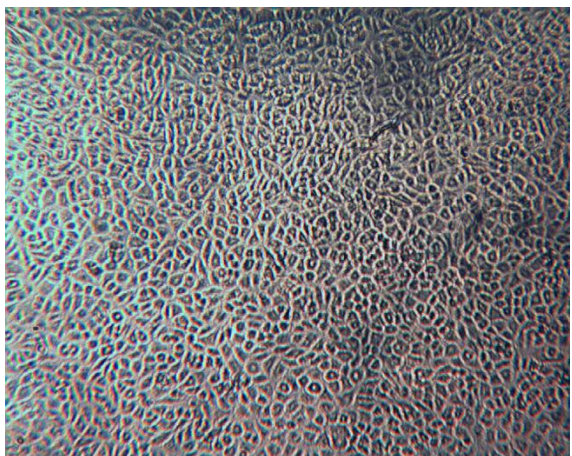
(e)

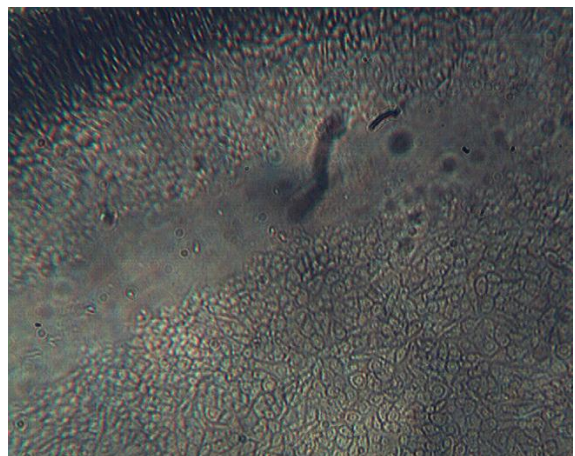
(f)

**Figure 10.** Cell morphology of L929 mouse fibroblasts cultured at different time points during 48-hour culture in sample extracts, DMEM as positive control and DMSO as negative control. (a, c): morphology of cells cultured in Collagen-Ph0.5 following 24 (a) and 48 (c) hours cell culture; (b, d): morphology of cells cultured in Collagen-Ph1.5 following 24 (b) and 48 (d) hours cell culture; (e, f): morphology of cells cultured in DMEM (e) and DMSO (f) following 48 hours cell culture.

This suggests that both types of sample extracts were non-toxic to L929 cells, so that cell confluence and metabolic activity were not decreased compared to cells cultured in cell culture media (positive control). Given that the samples tested identified the two composition extremes in terms of Ph feed ratio, both cell morphology and MTS results suggest that the sample extracts are well-tolerated by L929 cells, so that a non-toxic covalent network could be expected following the crosslinking reaction. In light of the observed sample biocompatibility, next steps will focus on *in vitro* investigations of these materials with specific cell types, including stem cells.

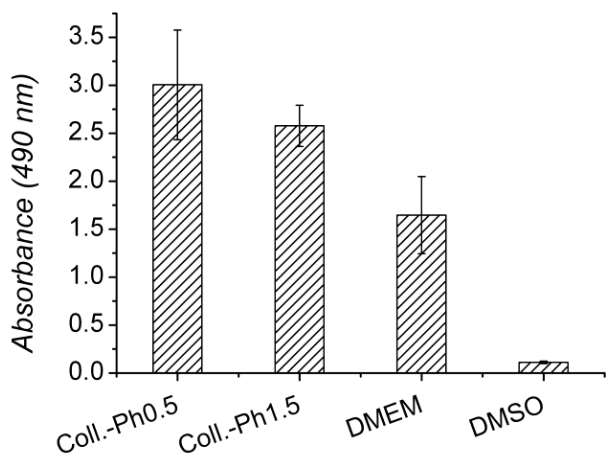

**Figure 11.** Formazan absorbance following MTS assay on L929 cells after 48-hour cell culture. Cells cultured on both Collagen-Ph0.5 and Collagen-Ph1.5 sample extracts showed higher absorbance compared to cells cultured in cell culture medium (DMEM) and dimethyl sulfoxide (DMSO), highlighting full hydrogel biocompatibility.

## 4. Conclusions

Triple-helical hydrogels were successfully formed via functionalisation of type I collagen with Ph as novel, bifunctional segment, and compared to state-of-the art EDC-crosslinked collagen. Resulting materials were characterised as for their protein conformation, network architecture, thermo-mechanical properties, degradability, bioactivity and cytotoxicity. Functionalisation with the aromatic segment provided the formation of a triple helical covalent network with tunable crosslinking density and swelling ratio, whereby enhanced macroscopic properties were obtained compared to systems with EDC-mediated intramolecular crosslinks. Consequently, resulting Ph-based systems displayed only minimal mass loss following 1-week incubation in either PBS or SBF, whilst nucleation of an amorphous calcium phosphate phase was initiated in retrieved samples. The presence of the Ph aromatic ring in resulting networks proved is likely to promote an interaction with calcium species present in solution, so that a specific site for calcium phosphate nucleation may be expected. L929 cell culture revealed no sign of cytotoxicity following 48-hour incubation in sample extract, whereby cell confluence was observed as in the case of culture in cell culture medium. Furthermore, the metabolic activity of cells cultured in the sample extracts was not decreased with respect to the one observed following cell culture in DMEM. Particularly based on their bioactivity and full biocompatibility, these materials have significant appeal for applications in mineralised tissue regeneration.

## Acknowledgements

This work was funded through WELMEC, a Centre of Excellence in Medical Engineering funded by the Wellcome Trust and EPSRC, under grant number WT 088908/Z/09/Z. The authors would like to thank Dr. P. Hine, J. Hudson, W. Vickers and S. Finlay, for kind assistance with WAXS, SEM/EDS, SBF preparation, and compression tests, respectively.